\documentclass[journal=jpclcd,manuscript=letter,layout=twocolumn]{achemso}

\usepackage[version=3]{mhchem} 
\usepackage{amssymb,amsmath}
\usepackage{graphicx}
\usepackage[usenames,dvipsnames,svgnames,table]{xcolor}
\usepackage[normalem]{ulem}
\usepackage{xr-hyper}
\usepackage{hyperref}
\usepackage{siunitx}
\usepackage{bm}
\usepackage{textcomp}
\usepackage{gensymb}
\usepackage{placeins}
\usepackage{physics}
\usepackage{nicefrac}
\usepackage[caption=false]{subfig}
\usepackage{xcolor} 
\usepackage{cuted}

\newcommand{\md}{\mathrm{d}}
\newcommand{\tDur}{t_{\rm prot}}
\newcommand{\tRelax}{t_{\rm relax}}
\newcommand{\torq}{\tau}
\newcommand{\Fo}{$\mathrm{F}_\mathrm{o}$\,\,}
\newcommand{\Fone}{$\mathrm{F}_1$\,\,}

\makeatletter
\newcommand*{\addFileDependency}[1]{
  \typeout{(#1)}
  \@addtofilelist{#1}
  \IfFileExists{#1}{}{\typeout{No file #1.}}
}
\makeatother

\newcommand*{\myexternaldocument}[1]{%
    \externaldocument{#1}%
    \addFileDependency{#1.tex}%
    \addFileDependency{#1.aux}%
}

\myexternaldocument{revised_SI}

\author{Deepak Gupta}
\affiliation{Department of Physics, Simon Fraser University, Burnaby, British Columbia V5A 1S6, Canada}
\altaffiliation{Institute for Theoretical Physics, Technical University of Berlin, Hardenbergstr. 36, D-10623 Berlin, Germany}
\author{Steven J.\ Large}
\altaffiliation{Current address: Viewpoint Investment Partners, Calgary, Alberta, Canada}
\affiliation{Department of Physics, Simon Fraser University, Burnaby, British Columbia V5A 1S6, Canada}
\author{Shoichi Toyabe}
\affiliation{Department of Applied Physics, Tohoku University, Aoba 6-6-05, Sendai, 980-8579, Japan}
\author{David A.\ Sivak}
\affiliation{Department of Physics, Simon Fraser University, Burnaby, British Columbia V5A 1S6, Canada}
\email{dsivak@sfu.ca} 

\title{Optimal Control of the F${_1}$-ATPase Molecular Motor}


\begin{document}

\begin{tocentry}
\includegraphics[width=\textwidth]{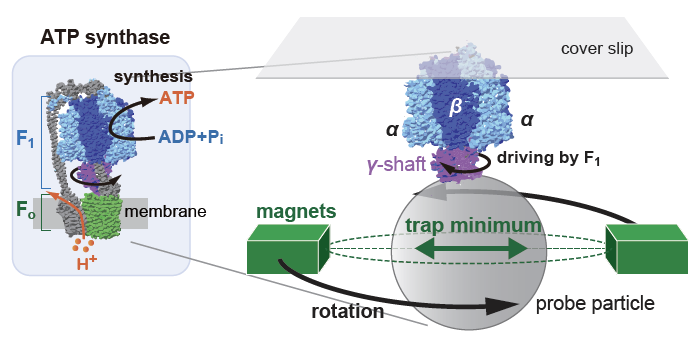}
\end{tocentry}

\begin{abstract}
F$_{1}$-ATPase is a rotary molecular motor that {\it in vivo} is subject to strong nonequilibrium driving forces. There is great interest in understanding the operational principles governing its high efficiency of free-energy transduction. Here we use a near-equilibrium framework to design a non-trivial control protocol to minimize dissipation in rotating \Fone to synthesize ATP. We find that the designed protocol requires much less work than a naive (constant-velocity) protocol across a wide range of protocol durations. Our analysis points to a possible mechanism for energetically efficient driving of \Fone \emph{in vivo} and provides insight into free-energy transduction for a broader class of biomolecular and synthetic machines.
\end{abstract}


Nanometer-sized biomolecular machines convert between different forms of free energy, while remaining in contact with a thermal environment. Consequently, fluctuations play a crucial role in a machine's dynamics~\cite{Seifert_ST,ST-2}; nevertheless, a machine achieves (on average) directed motion that is consistent with the second law of thermodynamics by transducing free energy, often stored in nonequilibrium chemical concentrations in the surrounding environment~\cite{ST-3}. It is of paramount interest to unravel the design principles governing effective free-energy transduction in biomolecular machines~\cite{Rev-1,Rev-2,Rev-3}.

${\rm F}_{\rm o}{\rm F}_1$-ATP synthase has attracted particular attention~\cite{boyer1993binding,FOF1-1,FOF1-2,FOF1-3}. This molecular motor is responsible for $\sim$95\% of the cellular synthesis of adenosine triphosphate (ATP)~\cite{yoshida2001atp,FOF1-4}. ATP production by the \Fone subunit is driven by rotation of its $\gamma$-shaft. \emph{In vivo}, the $\gamma$-shaft rotates by utilizing free energy from proton flux through the membrane-embedded \Fo subunit. The $\gamma$-shaft can rotate as fast as $\sim$350 revolutions per second~\cite{FOF1-2}, while maintaining high efficiency~\cite{Soga4960}. It is thus of significant interest to measure and quantitatively understand energy conversion during rapid mechanical driving of \Fone to synthesize ATP~\cite{Toyabe_2015,Frasch_1,Frasch_2}.

F$_{\rm o}$F$_1$ can operate in either direction~\cite{Huxley}: An excess of ATP drives counter-rotation of the $\gamma$-shaft and transports protons against their concentration difference. This reversibility is also observed in isolated F$_1$, which under sufficiently high torque synthesizes ATP~\cite{rondelez2005highly,itoh2004mechanically}, but in the absence of torque hydrolyzes ATP and counter-rotates the $\gamma$-shaft~\cite{noji1997direct,Huxley}. Experiments suggest that this `hydrolysis mode' proceeds via 120$^\circ$ rotations of the $\gamma$-shaft~\cite{yasuda2001resolution}, composed of two substeps: an $80^\circ$ step involving ATP binding, and a $40^\circ$ step involving ATP hydrolysis (catalysis)~\cite{Shimabukuro14731}. Overall the $\gamma$-shaft rotates 360$^\circ$ by hydrolyzing 3 ATPs.

Rapidly driving the $\gamma$-shaft by applying external torque inevitably produces dissipation, the difference between the work performed and the free energy transduced during synthesis or hydrolysis. It remains enigmatic how ATP synthase achieves highly efficient energetic conversion despite such rapid operation. In particular, what manner of rapid forced rotation with the $\gamma$-shaft achieves efficient energy transmission? 

Here we theoretically address efficient driving procedures (\emph{protocols}) to rapidly force rotation of the $\gamma$-shaft. A near-equilibrium framework~\cite{DS-PRL} has proven experimentally useful in designing a protocol for switching between folded and unfolded conformations of single DNA hairpins~\cite{Tafoya5920}, and similarly successful in simulations of barrier crossing~\cite{Barrier,Blaber_2022}, rotary motors~\cite{Joseph}, Ising models~\cite{RotskoffCrooks_PRE15,RotskoffEVE_PRE17,Miranda}, and other model systems~\cite{ZulkowskiDeWeese_PRE12,ZulkowskiDeWeese_PO13,erasure,BonancaDeffner_JCP14}. In this Letter we design a protocol that (near equilibrium) minimizes dissipation in experimentally accessible rotation of the $\gamma$-shaft driving \Fone to synthesize ATP. We find that the designed protocol outperforms the naive (constant-velocity) protocol for a considerable range of protocol durations, often far from equilibrium. Such protocols hint at how \Fo might rotate the $\gamma$-shaft in an efficient manner.

\begin{figure}[t!] 
    \centering
    \includegraphics[width = \columnwidth]{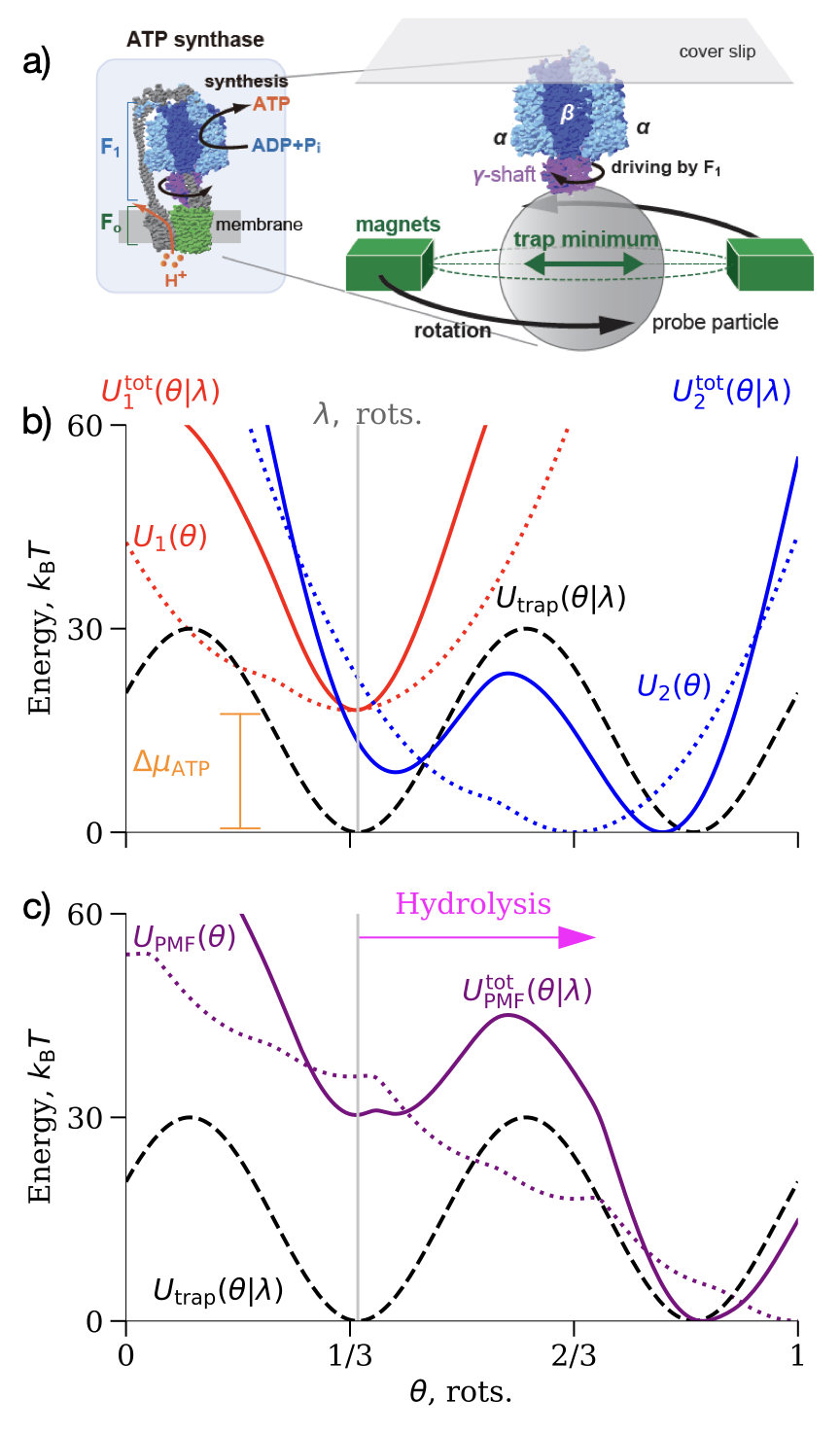}
    \caption{
    a) Schematic of modeled experiment. b) Effective potential $U_{1,2}(\theta)$ (dotted), trap potential $U_{\rm trap}(\theta|\lambda)$ (black dashed), and total potentials $U^{\rm tot}_{1,2}(\theta|\lambda) \equiv U_{1,2}(\theta)+U_{\rm trap}(\theta|\lambda)$ (solid). c) Potential of mean force $U_{\rm PMF}(\theta)$ (dotted), trap potential $U_{\rm trap}(\theta|\lambda)$ (black dashed), and total potential $U^{\rm tot}_{\rm PMF}(\theta|\lambda) \equiv U_{\rm PMF}(\theta)+U_{\rm trap}(\theta|\lambda)$ (solid). Each potential is plotted as a function of $\gamma$-shaft angle $\theta$, for fixed trap minimum $\lambda=0.41$ (vertical line), chemical drive $\Delta \mu_{\rm ATP}=18~k_{\rm B}T$, and trap strength $E^\ddag = 30~k_{\rm B}T/{\rm rad}^{2}$.}
    \label{fig:scheme}
\end{figure}

The totally asymmetric allosteric model (TASAM)~\cite{kawaguchi2014nonequilibrium} of \Fone describes the evolution of a rotational degree of freedom $\theta \in [0,2\pi]$ obeying periodic boundary conditions, corresponding to a bead attached to the $\gamma$-shaft (Fig.~\ref{fig:scheme}a provides a schematic of the modeled experiment). The TASAM is constructed to recover the steady-state and kinetic behavior of F$_1$ hydrolyzing ATP during single-molecule experiments~\cite{Toyabe_2012,Toyabe17951}.
The $\ell = 40^\circ$ step is modeled by the system switching between two harmonic potentials of the same spring constant $k\approx 20~k_{\rm B}T/{\rm rad}^{2}$, and with minima offset by the free-energy difference $\widetilde{\Delta\mu} = 5.2~k_{\rm B}T$ between the catalytic-dwell and binding-dwell states. Coarse-graining over the fast 40$^\circ$ step gives the effective potential 
\begin{align}
    \beta U_n(\theta) \equiv -\ln\big[e^{-\frac{1}{2}\beta k(\theta+\ell-n \xi)^2-\beta\widetilde{\Delta\mu}}+e^{-\frac{1}{2}\beta k(\theta-n \xi)^2}\big] \ ,\label{un-pot}
\end{align}
for $n=0,\pm 1,\pm 2,\dots$. $\beta\equiv 1/(k_{\rm B}T)$ is the inverse temperature. The first and second terms in square brackets respectively indicate the catalytic-dwell and binding-dwell states. The $80^{\circ}$ step is modeled by a hop between adjacent effective potentials $U_{n}(\theta)$ and $U_{{n}\pm1}(\theta)$ angularly separated by $\xi = 2\pi/3~{\rm rad}$ [see $U_{1,2}(\theta)$ in Fig.~\ref{fig:scheme}b].

Thus, the overall hydrolysis/synthesis of one ATP molecule is modeled by switching the effective potential, $U_{n}(\theta) \to U_{{n}\pm1}(\theta)$, where the `+' (`-') sign is for hydrolysis (synthesis). The corresponding transition rates obey the local detailed-balance condition~\cite{kawaguchi2014nonequilibrium}
\begin{align}
    \dfrac{R^+_n(\theta)}{R^-_{n+1}(\theta)} =e^{\beta[U_n(\theta)-U_{n+1}(\theta)+\Delta \mu_{\rm ATP}]} \ , \label{ldb}
\end{align}
for chemical-potential difference (hereafter \emph{chemical drive}) $\Delta \mu_{\rm ATP}\geq 0$ due to synthesis of one ATP, favoring ATP hydrolysis and $\gamma$-shaft counter-rotation. The potential switches with respective forward and backward transition rates
\begin{subequations}
\label{W-eqn}
\begin{align}
    R_{n}^+(\theta) &= \Gamma \label{fwdR},\\
    R_{n+1}^-(\theta) &= \Gamma e^{-\beta [U_n(\theta)-U_{n+1}(\theta)+\Delta \mu_{\rm ATP}]}\label{bwdR} \ .
\end{align}
\end{subequations}
$\Gamma$ is a rate constant characterizing the chemical reaction (Supporting Information~\cite{SM} and Fig.~\ref{experimental_data} relate $\Gamma$ and [ATP] at fixed $\Delta \mu_{\rm ATP}$).  Among the range of possible splittings of angular dependence between the forward and backward transition rates that satisfy local detailed-balance~\eqref{ldb}, the experimental kinetics in Refs.~\cite{Toyabe17951,Toyabe_2012} are best fit by this splitting~\cite{kawaguchi2014nonequilibrium}.

Experimentally, \Fone is driven by confining the magnetic bead attached to the $\gamma$-shaft in a magnetic trap~\cite{itoh2004mechanically, rondelez2005highly, Saita2015,Palanisami2010}, whose minimum is dynamically rotated. Modeling such a magnetic trap, here we consider a sinusoidal trap potential (see Fig.~\ref{fig:scheme}),
\begin{align}
    U_{\rm trap}(\theta|\lambda) \equiv -\tfrac{1}{2}E^\ddag \cos 2(\theta-2\pi\lambda) \ , \label{trap}
\end{align}
with time-dependent control parameter $\lambda \in [0,1]$ determining the angle of the two minima differing by 180$^\circ$ and separated by barriers of height $E^\ddag$ (parametrizing trap strength).

Thus, subject to this external trap, the $\gamma$-shaft angle $\theta$ dynamically evolves according to
\begin{align}
    \dot \theta = -\beta D\, \partial_\theta 
    U_n^{\rm tot}(\theta|\lambda) + \sqrt{2D}~\eta(t) \label{W-dynamics-main} \ ,
\end{align}
where the dot indicates a time derivative, $U_n^{\rm tot}(\theta|\lambda) \equiv U_n(\theta)+U_{\rm trap}(\theta|\lambda)$ is the total potential, $D = 13.7~{\rm rad}^2/{\rm s}$ is the rotational diffusion constant~\cite{Toyabe_2012}, and $\eta(t)$ is Gaussian white noise with zero mean and unit variance. For $\Delta \mu_{\rm ATP} \neq 0$, the system eventually reaches a nonequilibrium steady state~\cite{risken1996fokker}. Chemical transition rates~\eqref{W-eqn} and mechanical dynamics~\eqref{W-dynamics-main}, subject to potentials~\eqref{un-pot} and \eqref{trap}, constitute the TASAM.

In the limit of fast switching between effective potentials when chemistry is fast compared to mechanics ($\Gamma \to \infty$, e.g., at high [ATP])~\cite{kawaguchi2014nonequilibrium}, the dynamics~\eqref{W-dynamics-main} reduce to
\begin{align}
 \dot \theta = -\beta D\,\partial_\theta U^{\rm tot}_{\rm PMF}(\theta|\lambda)+\sqrt{2D}\, \eta(t) \ ,\label{tot-lan-dyn}
\end{align}
for total potential energy $U^{\rm tot}_{\rm PMF}(\theta|\lambda) \equiv U_{\rm PMF}(\theta)+U_{\rm trap}(\theta|\lambda)$ and potential of mean force, averaging over all effective potentials (see Fig.~\ref{fig:scheme}c):
\begin{align}
    \beta U_{\rm PMF}(\theta) \equiv -\ln \sum_{n=-\infty}^{+\infty} e^{-\beta [U_{n}(\theta)-n\Delta \mu_{\rm ATP}]} \ .\label{pmf-pot}
\end{align}

We seek a driving protocol that minimizes dissipation while rotating the $\gamma$-shaft (for theoretical developments when control is much finer-grained, see Refs.~\cite{NakazatoIto_PRR21,Ito_arXiv22}). For a harmonically confined Brownian particle, minimum-dissipation protocols have been analytically solved for arbitrary protocol duration~\cite{OC_PRL,OC-2}. For more complicated scenarios such as this model, no analytical solution is known; nevertheless, linear-response theory gives an approximately dissipation-minimizing protocol~\cite{DS-PRL}. Up to the linear-response approximation, the instantaneous excess power (that exceeding the quasistatic power) during dynamic variation of control parameter $\lambda$ is
\begin{align}
    P_{\rm ex}(t) \approx \zeta(\lambda) \bigg(\dfrac{\md\lambda}{\md t}\bigg)^2\ .\label{ex-pow}
\end{align}
Its time integral over protocol duration $\tDur$ gives the excess work $W_{\rm ex}\equiv W -\Delta F = \int_0^{\tDur}~{\md}t~P_{\rm ex}(t)$, for protocol work $W$ and equilibrium free-energy change $\Delta F$ from initial to final control-parameter values. $\zeta(\lambda)$ is a generalized friction coefficient obtained here by integrating the equilibrium torque autocovariance:
\begin{align}
    \zeta(\lambda) \equiv \beta \int_0^\infty~{\md}t~\langle \delta \torq(0)~\delta \torq(t) \rangle_{\lambda} \ . \label{fric}
\end{align}
Angle brackets $\langle \dots \rangle_\lambda$ indicate a steady-state average at fixed $\lambda$. $\delta \torq(t) \equiv \torq(t) - \langle \torq\rangle_{\lambda}$ is the deviation of the conjugate torque $\torq \equiv -\partial_\lambda U_{\rm trap}(\theta|\lambda) = 2\pi E^\ddag \sin 2(\theta-2\pi\lambda)$ from its equilibrium average. 
The generalized friction coefficient can be decomposed as
\begin{align}
   \zeta(\lambda) = \beta \langle (\delta \torq)^2 \rangle_{\lambda}~\tRelax(\lambda) \ ,
\end{align}
the product of the torque variance $\langle (\delta \torq)^2 \rangle_{\lambda}$ and the torque relaxation time
\begin{align}
\tRelax(\lambda) \equiv \int_0^\infty~{\md}t~\dfrac{\langle \delta \torq(0)~\delta \torq(t) \rangle_{\lambda} }{\langle (\delta \torq)^2 \rangle_{\lambda}} \ . \label{tau-relax}
\end{align}
In the linear-response regime, the minimum-dissipation protocol proceeds with velocity inversely proportional to the square root of the generalized friction coefficient~\cite{DS-PRL}:
\begin{align}
   \dfrac{{\md}\lambda^{\rm des}}{{\md}t} \propto [\zeta(\lambda)]^{-1/2} \ .\label{des-prot} 
\end{align}
Dissipation is reduced by driving slower where system resistance is greatest (due to large fluctuations and/or slow relaxation) and compensating by driving faster where system resistance is least. Imposing boundary conditions $\lambda(0)=\lambda_{\rm i}$ and $\lambda(\tDur) = \lambda_{\rm f}$ fixes the proportionality constant. Such a designed protocol $\lambda^{\rm des}(t)$ gives (up to linear response) constant excess power.

\begin{figure} 
    \centering
    \includegraphics[width=\columnwidth]{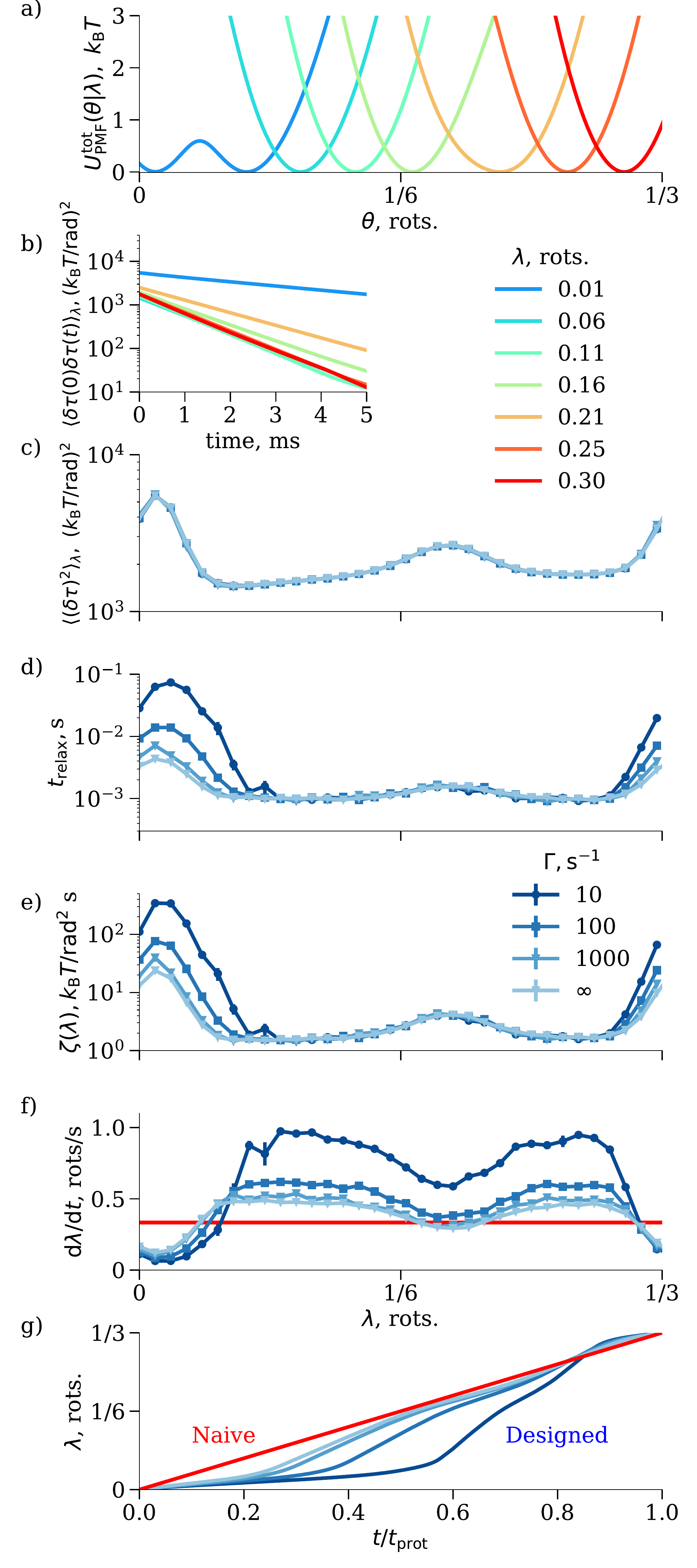}
    \caption{(a) Total potential energy as a function of $\gamma$-shaft angle $\theta$. (b) Torque autocovariance as a function of time. (c) Torque variance, (d) torque relaxation time, (e) friction coefficient, and (f) protocol velocity, each as a function of trap minimum $\lambda$. (g) Protocol as a function of time. (f,g) Red lines: naive (constant-velocity) protocols; points/curves: designed protocols. Potential switching rate $\Gamma = \infty$ (a,b), and $10/100/1000/\infty \, {\rm s}^{-1}$ (c-g). Throughout, trap strength $E^\ddag =30~k_{\rm B}T/{\rm rad}^2$ and chemical drive $\Delta\mu_{\rm ATP}=18~k_{\rm B}T$. In all figures, error bars show one standard error of the mean.} 
    \label{fig:fric-var-relax}
\end{figure}

Figure~\ref{fig:fric-var-relax}a shows $U^{\rm tot}_{\rm PMF}(\theta|\lambda)$ for different trap minima $\lambda$ and fast switching. For some $\lambda$'s, the total potential has two metastable states, with a small ($<0.5~k_{\rm B}T$) barrier.

The friction coefficient~\eqref{fric} is obtained through measurement of the equilibrium torque autocovariance for a stationary trap (i.e., fixed $\lambda$) and hence fixed total potential-energy landscape (Fig.~\ref{fig:fric-var-relax}a). Figure~\ref{fig:fric-var-relax}b shows this torque autocovariance function. For those trap minima $\lambda$ giving two metastable states (Fig.~\ref{fig:fric-var-relax}a), the torque autocovariance relaxes particularly slowly. 

\begin{figure}[t!] 
    \centering
    \includegraphics[width = \columnwidth]{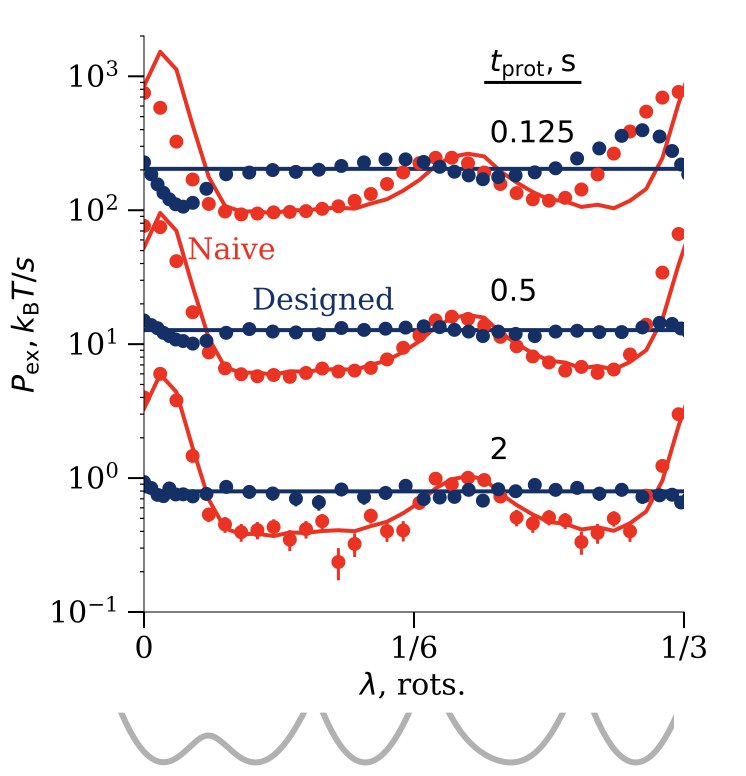}
    \caption{Steady-state excess power while driving ATP synthesis, as a function of trap minimum $\lambda$, for protocol durations $\tDur = 0.125,~0.5$, and 2 s, for naive (red) and designed (blue) protocols (Fig.~\ref{fig:fric-var-relax}g) in the full model (points) and the linear-response approximation~\eqref{ex-pow} (curves). Throughout, trap strength $E^\ddag = 30~k_{\rm B}T/{\rm rad}^2$, chemical drive $\Delta \mu_{\rm ATP}=18~k_{\rm B}T$, and potential switching rate $\Gamma \to \infty$. Gray curves at bottom depict the total potential at the corresponding trap minimum.}
    \label{fig:ex-pow-main}
\end{figure}

Figure~\ref{fig:fric-var-relax}c shows the torque variance computed from the equilibrium torque fluctuations. The torque variance is independent of $\Gamma$, as expected since varying $\Gamma$ modifies system relaxation between landscapes but not the landscapes themselves. Similarly, Fig.~\ref{fig:fric-var-relax}d shows the torque relaxation time~\eqref{tau-relax}, $\tRelax(\lambda)$. The product of the torque variance and torque relaxation time gives the generalized friction coefficient (Fig.~\ref{fig:fric-var-relax}e). The peak relaxation time (generally for trap minimum $\lambda$ such that the system has significant probability in multiple potentials) is higher for smaller $\Gamma$, which slows system relaxation between distinct potentials. The torque variance, torque relaxation time, and friction coefficient each have a local maximum for the trap minimum near the shoulder of $U_n(\theta)$ (Fig.~\ref{fig:scheme}b) and $U_{\rm PMF}(\theta)$ (Fig.~\ref{fig:scheme}c). As expected, these local maxima in friction coefficient coincide with the experimental local minima of F$_1$’s rotational velocity during hydrolysis~\cite{Frasch_1}.

According to \eqref{des-prot}, excess work is reduced by slowing down where friction is high, giving more time for thermal fluctuations to overcome the energy barrier separating metastable states (Fig.~\ref{fig:fric-var-relax}a) and thereby reducing necessary work. Figure~\ref{fig:fric-var-relax}f compares the control-parameter velocities of designed protocols and naive ones (i.e., with constant velocity). Integrating the control-parameter velocity gives the protocol as a function of time (Fig.~\ref{fig:fric-var-relax}g). 

To compute excess power and work, we numerically simulate (without assuming linear response) the full model of the system, Eqs.~\ref{W-dynamics-main} and \ref{tot-lan-dyn}, driven by naive and designed protocols across a full rotation driving ATP synthesis (decreasing $\lambda$). 

Figure~\ref{fig:ex-pow-main} shows the steady-state excess power of the full model for fast potential switching rate ($\Gamma\to\infty$), with the linear-response approximation~\eqref{ex-pow} as a reference. The linear-response approximation is quite accurate in the long-duration limit since the system remains sufficiently close to equilibrium. For the naive protocol, excess power is strongly peaked where the $\gamma$-shaft experiences high resistance (Fig.~\ref{fig:fric-var-relax}e). The designed protocol compensates by driving slower there (Fig.~\ref{fig:fric-var-relax}d), thereby flattening the excess power across the protocol. 

\begin{figure}[t!]
    \centering
    \includegraphics[width = \columnwidth]{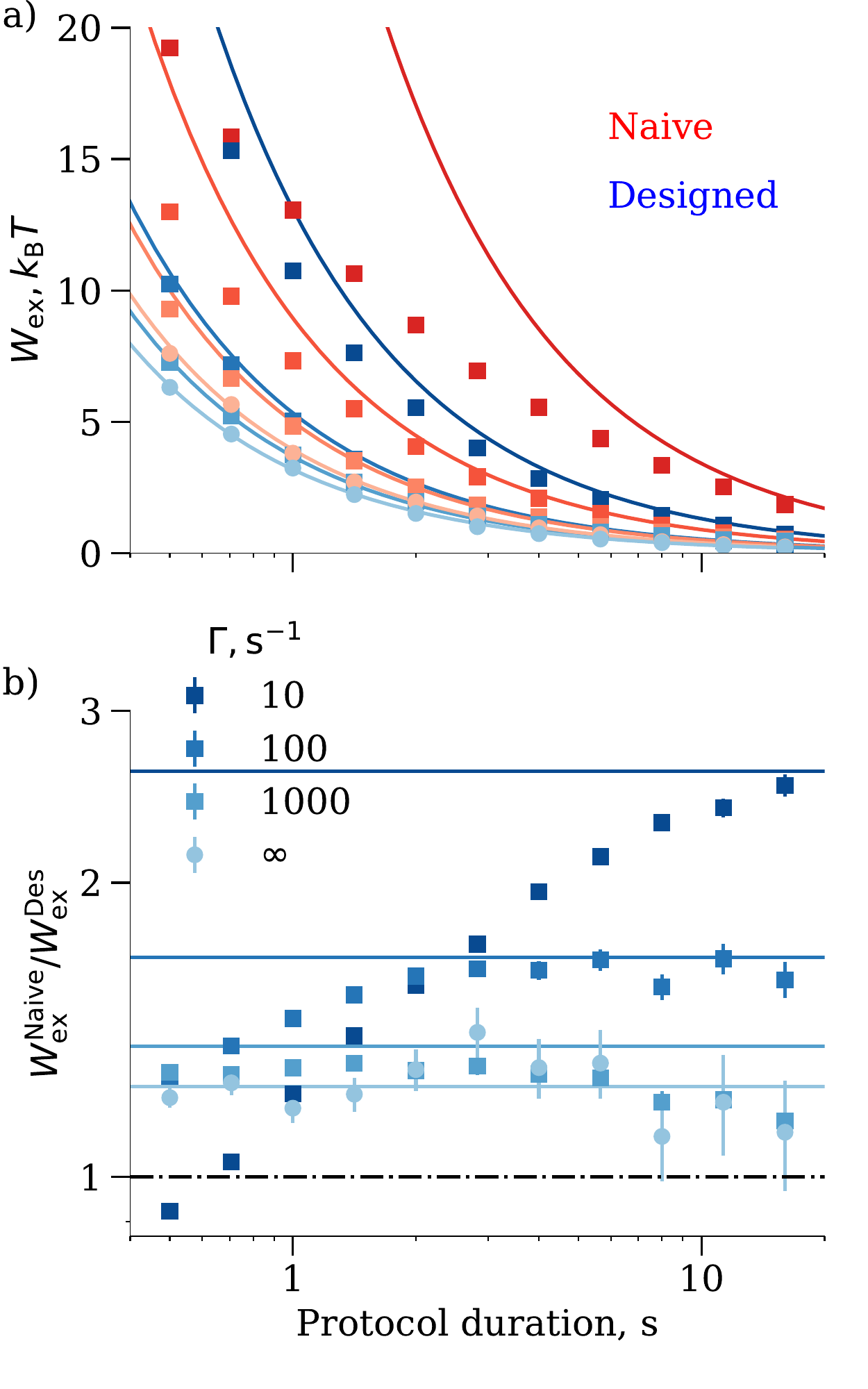}
    \caption{Excess work while driving ATP synthesis. a) Naive (red) and designed (blue) excess works. b) Ratio of naive and designed excess works. Points show the full model and curves the linear-response approximation. Color intensity decreases with switching rate $\Gamma$. Throughout, trap strength $E^\ddag = 30~k_{\rm B}T/{\rm rad}^2$ and chemical drive $\Delta \mu_{\rm ATP} = 18~k_{\rm B}T$.}
    \label{fig:W-fig-work}
\end{figure}

Finally, Figure~\ref{fig:W-fig-work}a shows excess work during naive and designed protocols driving ATP synthesis, for different switching rates $\Gamma$. 
As expected, the work decreases with protocol duration and with $\Gamma$. For longer durations, the linear-response approximation is quite accurate. 

Figure~\ref{fig:W-fig-work}b shows the ratio of excess work during naive and designed protocols. For the slowest protocols and slowest switching rate, designed protocols reduce excess work $\sim 2.5\times$ compared to naive. At longer protocol duration, the excess-work ratio for the slowest switching rate is almost twice that of the fastest switching rate ($\Gamma\to \infty$). A slower switching rate produces greater variation in the friction coefficient (Fig.~\ref{fig:fric-var-relax}e), leading to designed protocols that differ more from naive ones (Fig.~\ref{fig:fric-var-relax}g) and thus offering greater advantage to using designed protocols.

Analogous driving of ATP hydrolysis (instead of synthesis) and different trap strengths and chemical drives produce qualitatively similar generalized friction coefficients and designed protocols (Fig.~\ref{fig:friction-SI}), excess power along the protocol (Fig.~\ref{fig:ex-pow-SI}), and excess work (Figs.~\ref{fig:work-hydro-W} and \ref{fig:work-ratio-SI}).

We have studied nonequilibrium driving of an experimentally motivated model~\cite{kawaguchi2014nonequilibrium,Toyabe_2012} of the F${_1}$-ATPase molecular motor. We found that the designed protocol (that near equilibrium minimizes work) indeed significantly decreases work for a wide range of protocol durations, 
including some driving F$_1$ far from equilibrium and dissipating as much as several $k_{\rm B}T$ per ATP synthesized. This demonstrates the utility of this linear-response framework significantly beyond its regime of strict validity.
This protocol allocates more time to regions where fluctuations are large and/or relaxation is slow 
(and where \Fone rotates slowest when hydrolyzing ATP~\cite{Frasch_1}),
and hence where greater time for system relaxation can make the most difference. In this model this is where the total potential is bimodal, due to \Fone fluctuating between distinct chemical states with correspondingly different mechanics.

Recent work has derived optimal protocols in the opposite limit of very fast driving~\cite{Blaber_PRE21}; these rapid protocols, and interpolations between them and near-equilibrium ones, would allow systematic characterization of efficiency across driving speeds. Our simple near-equilibrium theory may also be helpful in physically constraining machine-learning algorithms~\cite{EngelBrenner_arXiv22} or in more direct numerical calculation~\cite{ZhongDeWeese_arXiv22} of optimal protocols far from equilibrium. Moreover, there is promise in exploring near-optimal protocols~\cite{GingrichGeissler_PNAS16}, analysis of which significantly broadens the class of successful control strategies. 

Our research opens avenues for future experimental investigation with an analogous setup to Refs.~\cite{Saita2015,Palanisami2010}. Such an experiment can be used to design and implement a rotational driving protocol employing our methodology. Our study predicts experimental conditions where such designs will produce substantial energy savings, informing experiments to either drive \Fone to efficiently produce ATP or efficiently harness mechanical energy from ATP hydrolysis.

The ${\rm F}_{\rm o}{\rm F}_1$ molecular motor transduces energy at high efficiency~\cite{silverstein2014exploration}. Our designed protocol may mimic an operational mechanism by which ${\rm F}_{\rm o}$ rotates F$_1$'s $\gamma$-shaft to synthesize ATP, driving slower where friction is higher, which we predict would save energy compared to constant-velocity driving. It would be interesting to probe the actual stochastic dynamics by which \Fo mechanically drives ${\rm F}_1$ {\em in vivo} and understand its correspondence to an effective nonequilibrium driving protocol. E.g., \Fo could approximate such a protocol if its metastable rotational states are out of phase with those of ${\rm F}_1$. More generally, our methodology may be applied to identify design principles for nanoscale free-energy transduction within and between other molecular machines.

\begin{acknowledgement}
We thank Alex Tong and Carlos Bustamante (Berkeley Physics) for enlightening discussions, and Adrianne Zhong (Berkeley Physics) for feedback on the manuscript. This work was supported by a Natural Sciences and Engineering Research Council of Canada (NSERC) Discovery Grant (D.A.S.), a Tier-II Canada Research Chair (D.A.S.), and the Deutsche Forschungsgemeinschaft (DFG, German Research Foundation) - Projekt-nummer 163436311 - SFB 910 (D.G.), and was enabled in part by support provided by BC DRI Group 
and the Digital Research Alliance of Canada (www.alliancecan.ca).
\end{acknowledgement}

\begin{suppinfo}
Supporting Information contains further numerical simulation results and details of simulation methods.
\end{suppinfo}

\clearpage
\appendix

\setcounter{equation}{0}
\setcounter{figure}{0}
\setcounter{table}{0}
\setcounter{page}{1}
\setcounter{section}{0}
\setcounter{subsection}{0}
\makeatletter
\renewcommand{\theequation}{S\arabic{equation}}
\renewcommand{\thefigure}{S\arabic{figure}}
\renewcommand{\thesection}{S\Roman{section}} 
\renewcommand{\thepage}{S\arabic{page}}

\begin{strip}
\section{\huge Supporting Information}
\end{strip}

\section{Relating $\Gamma$ to [ATP]}
\begin{figure*}
    \centering
    \includegraphics[width = \textwidth]{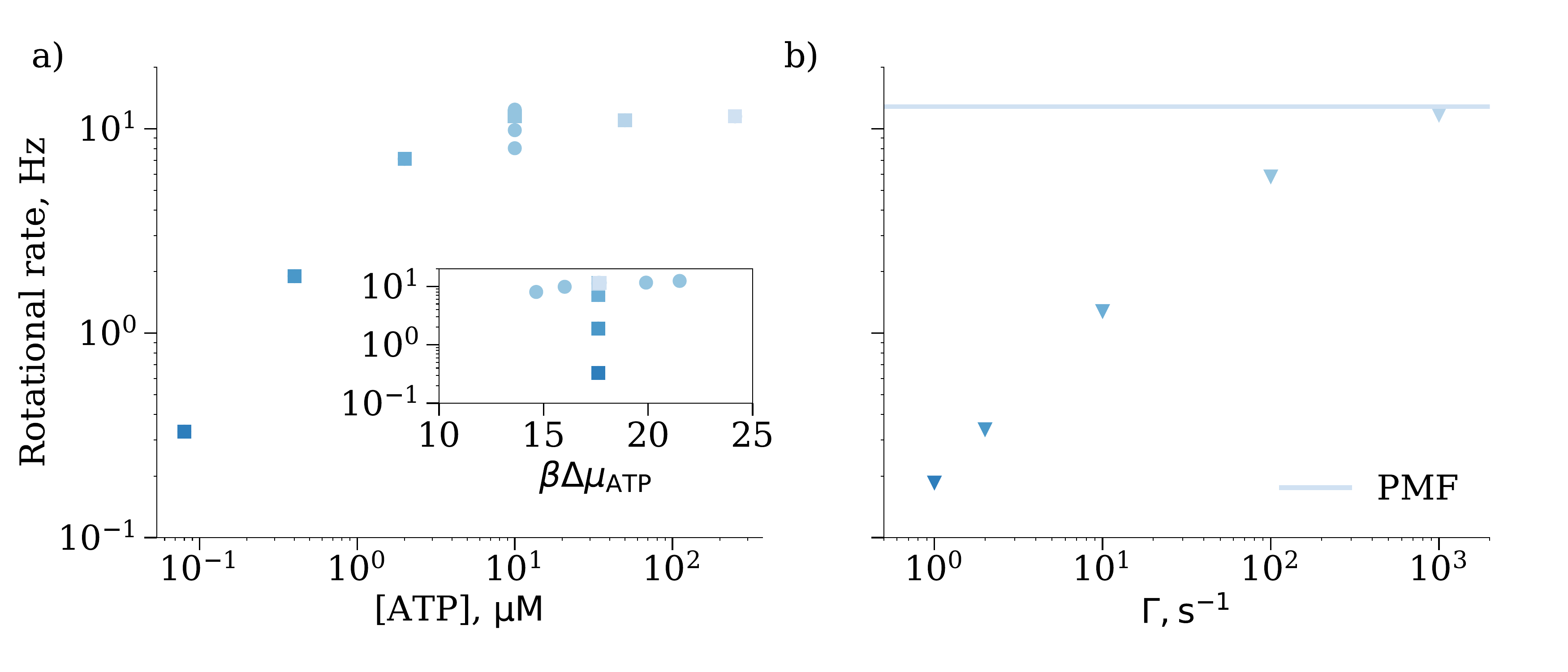}
    \caption{$\gamma$-shaft rotation rate (a) in experiment~\cite{Toyabe17951} as a function of [ATP] and of chemical drive $\beta \Delta \mu_{\rm ATP}$ (inset), and (b) in simulation as a function of switching rate $\Gamma$ for slow-switching dynamics~\eqref{W-dynamics-main} (triangles) or fast-switching dynamics~\eqref{tot-lan-dyn} (horizontal line). Color intensity decreases with [ATP] for chemical drive $\Delta \mu_{\rm ATP} \approx 17.6~k_{\rm B}T$.}
    \label{experimental_data}
\end{figure*}

Figure~\ref{experimental_data}a shows experimental rotational rates from~\cite{Toyabe17951}: The rotation rate strongly increases with [ATP] and weakly increases with $\Delta \mu_{\rm ATP}$. Figure~\ref{experimental_data}b shows numerical Langevin simulations of the rotation rate as a function of $\Gamma$: The rotation rate increases with $\Gamma$ and asymptotes to the rotation rate for fast-switching dynamics~\eqref{tot-lan-dyn} ($\Gamma \to \infty$). At fixed $\Delta \mu_{\rm ATP}$, comparing Figs.~\ref{experimental_data}a and b permits calibration of switching rate $\Gamma$: $\Gamma$ ranging from 1 to 1,000 Hz reproduces rotational rates for [ATP] ranging from 0.1 to 100 $\mu$M. 

\section{Torque fluctuations and designed protocol for $\Gamma \to \infty$: varying $E^{\ddag}$ and $\Delta \mu_{\rm ATP}$}
\begin{figure}[htbp]
    \centering
    \includegraphics[width = \columnwidth]{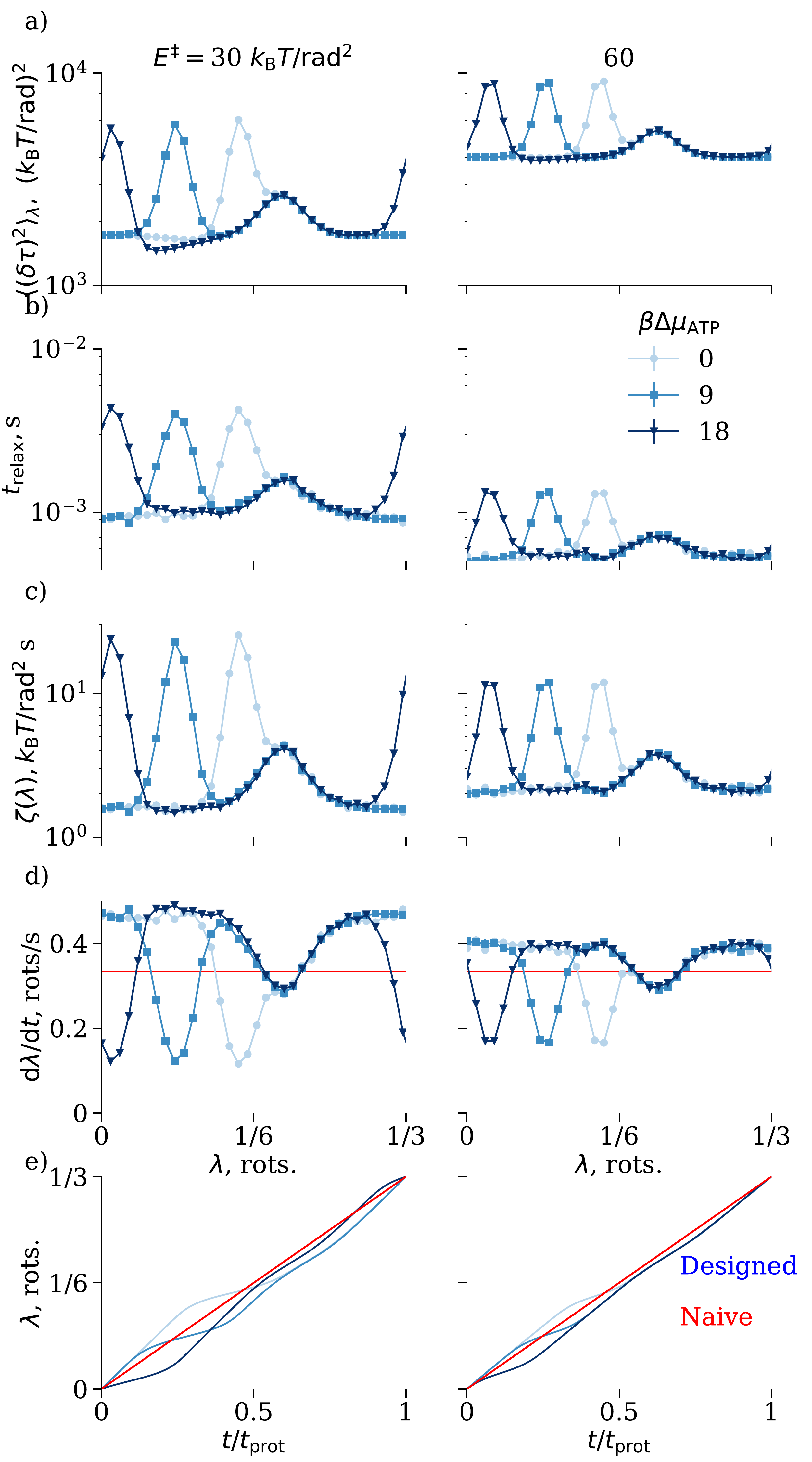}
    \caption{Equilibrium torque fluctuations and designed protocol. (a) Torque variance, (b) torque relaxation time, (c) friction coefficient, and (d) protocol velocity, each  as a function of trap minimum, $\lambda$. (e) Protocols as a function of time. (d,e) Red lines indicate naive protocols and curves/points indicate designed protocols. Trap strengths $E^\ddag = 30$ (left) and $60~k_{\rm B}T/{\rm rad^2}$ (right). Color intensity increases with chemical drive $\Delta\mu_{\rm ATP}$.}
    \label{fig:friction-SI}
\end{figure}

Figures~\ref{fig:friction-SI}a-d respectively show the torque variance, torque relaxation time, friction coefficient, and rate of change of protocol, each as a function of trap minimum $\lambda$, for two trap strengths $E^\ddag$ and three chemical drives $\Delta \mu_{\rm ATP}$. Figure~\ref{fig:friction-SI}e shows the naive and designed protocols obtained by integrating their respective protocol velocities (Fig.~\ref{fig:friction-SI}d)

The torque variance empirically increases with trap strength $E^\ddag$; this is intuitive, as in a simple quadratic potential the equipartition theorem~\cite{esm} dictates that the torque variance is directly proportional to $E^\ddag$. The torque variance, torque relaxation time, and friction coefficient each are globally maximized at the trap minimum $\lambda$ where $U^{\rm tot}_{\rm PMF}(\theta|\lambda)$ has two metastable states (see the leftmost potential in Fig.~\ref{fig:fric-var-relax}a), and have a local maximum for $\lambda$ at the PMF's shoulder (Fig.~\ref{fig:scheme}c). The friction coefficient's maximum value decreases with trap strength $E^\ddag$, because the barrier height in the double-well total potential $U^{\rm tot}_{\rm PMF}(\theta|\lambda)$ decreases with $E^\ddag$. Further, the global maximum shifts leftwards as $\Delta \mu_{\rm ATP}$ increases since this reduces the $\lambda$ at which the total potential $U^{\rm tot}_{\rm PMF}$ has two metastable states, and the insensitivity of $U_{\rm PMF}$'s shoulder to variation of $\Delta \mu_{\rm ATP}$ reflects that the local maximum is similarly insensitive. 

\section{Excess power for $\Gamma \to \infty$: varying direction, $E^{\ddag}$, and $\Delta \mu_{\rm ATP}$}
\begin{figure*}[t!]
    \centering
    \includegraphics[width=\textwidth]{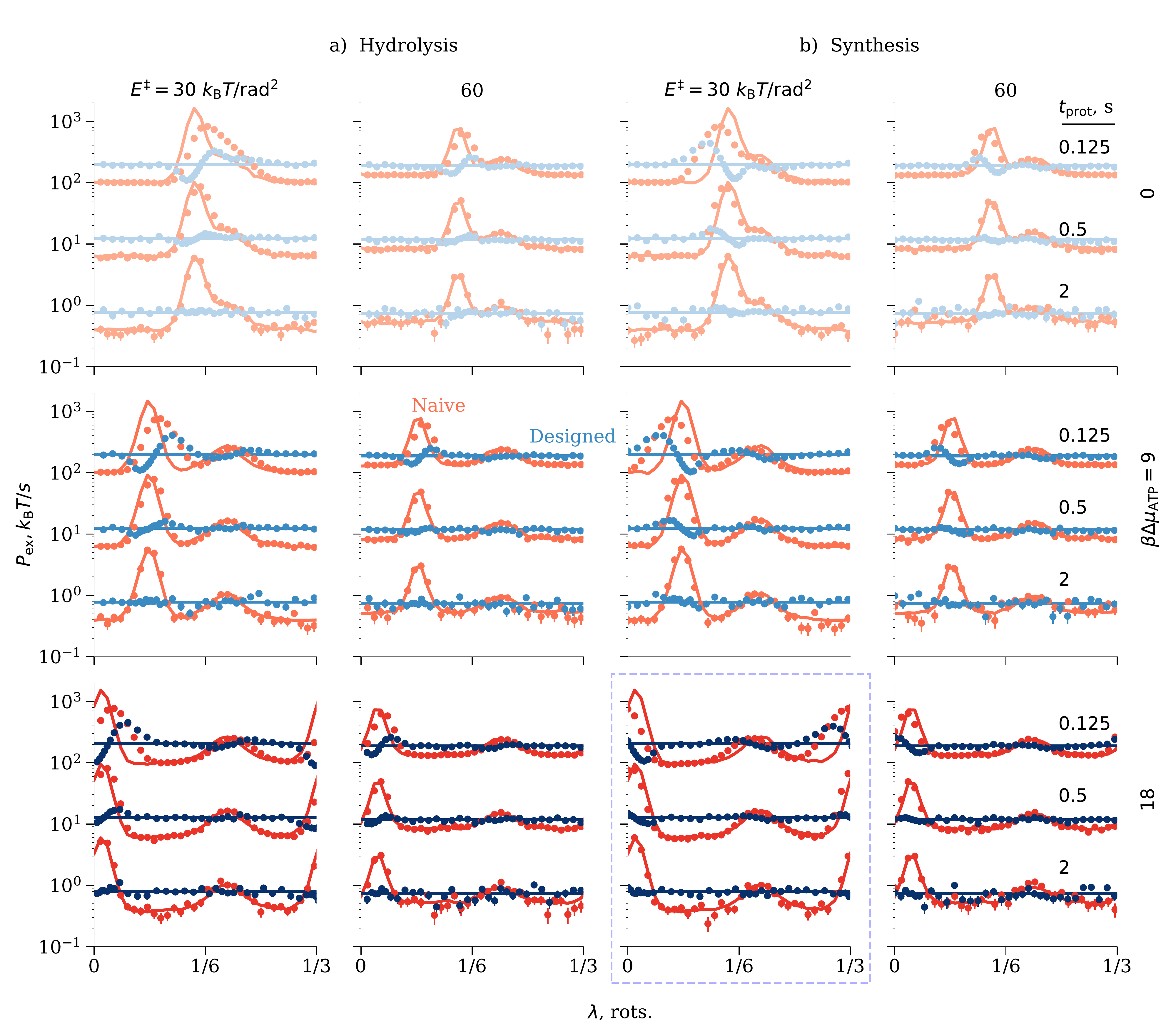} 
    \caption{Steady-state excess power as a function of $\lambda$, during driven (a) hydrolysis (increasing $\lambda$) and (b) synthesis (decreasing $\lambda$), naive (red) and designed (blue) protocols, for the full model (points) and the linear-response approximation~\eqref{ex-pow}. Blue dashed box indicates data shown in Fig.~\ref{fig:ex-pow-main}. Trap strengths $E^\ddag = 30$ and $60~k_{\rm B}T/{\rm rad}^2$, chemical drives $ \Delta\mu_{\rm ATP} = 0, 9,$ and $18~k_{\rm B}T$ (top to bottom), and protocol durations $\tDur = 0.125, 0.5$, and $2$~s.}
    \label{fig:ex-pow-SI}
\end{figure*}

Figure~\ref{fig:ex-pow-SI} compares the steady-state excess power $P_{\rm ex}$ as a function of trap minimum $\lambda$, during driven hydrolysis (increasing $\lambda$) and synthesis (decreasing $\lambda$), for three protocol durations $\tDur$, two trap strengths $E^{\ddag}$, and three chemical drives $\Delta \mu_{\rm ATP}$. The excess power is qualitatively similar for the two trap strengths, and lower for higher trap strength. Since during hydrolysis the $\gamma$-shaft rotates in the opposite manner than during synthesis, the excess power for the full model flips. In contrast, the linear-response approximation is the same for driven synthesis and hydrolysis. For longer durations, the linear-response approximation~\eqref{ex-pow} closely matches the full model, as expected. For intermediate duration, agreement is poor near the peak of the friction coefficient (where linear response predicts that excess power peaks under naive protocols). Agreement is better for a stronger trap since the system mostly follows the trap for the entire duration. 

\section{Excess work driving ATP hydrolysis}
\begin{figure}
    \centering
    \includegraphics[width = \columnwidth]{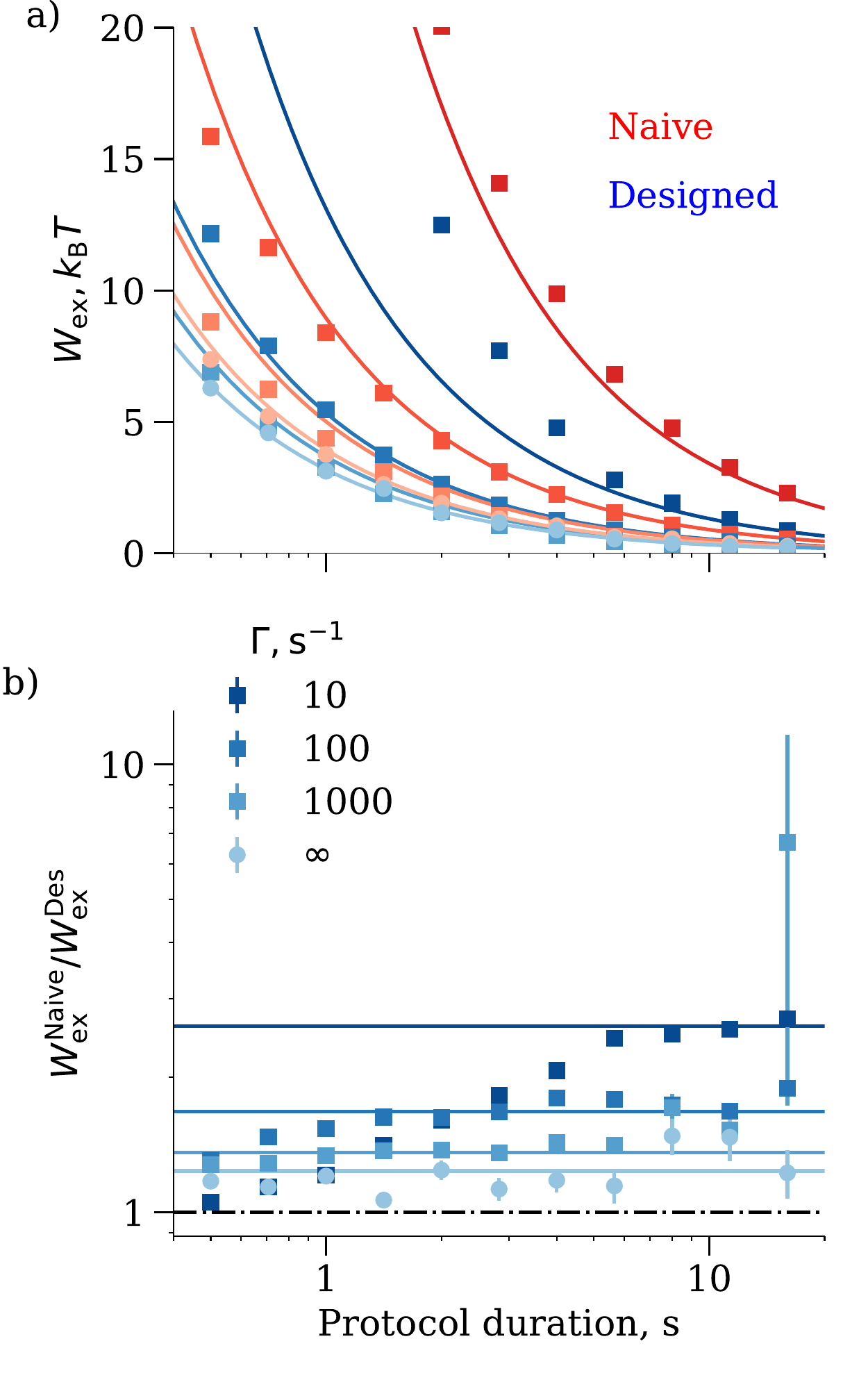}
    \caption{Excess work when driving hydrolysis. (a) Excess work for naive (red) and designed (blue) protocols, and (b) ratio of naive and designed excess works, each as a function of protocol duration, for the full model (points) and the linear-response approximation (curves). Trap strength $E^\ddag = 30~k_{\rm B}T/{\rm rad}^2$ and chemical drive $\Delta \mu_{\rm ATP} = 18~k_{\rm B}T$. Color intensity decreases with switching rate $\Gamma$. In all figures, error bars show one standard error of the mean.}
    \label{fig:work-hydro-W}
\end{figure}

Figure~\ref{fig:work-hydro-W} shows the excess works by rotating the $\gamma$-shaft employing the naive and designed protocols (see Fig.~\ref{fig:fric-var-relax}g) to drive ATP hydrolysis, for different values of switching rate $\Gamma$. The linear-response approximation is quite accurate in the long-protocol duration limit. Qualitatively, the excess-work features during hydrolysis are similar to those during synthesis, and the ratio of naive to designed excess work for our slowest switching rate ($\Gamma=10~{\rm s}^{-1}$) is almost double that in the limit $\Gamma\to \infty$.

\section{Excess work for $\Gamma \to \infty$: varying direction, $E^{\ddag}$, and $\Delta \mu_{\rm ATP}$}
\begin{figure*}[t!]
    \centering
    \includegraphics[width = \textwidth]{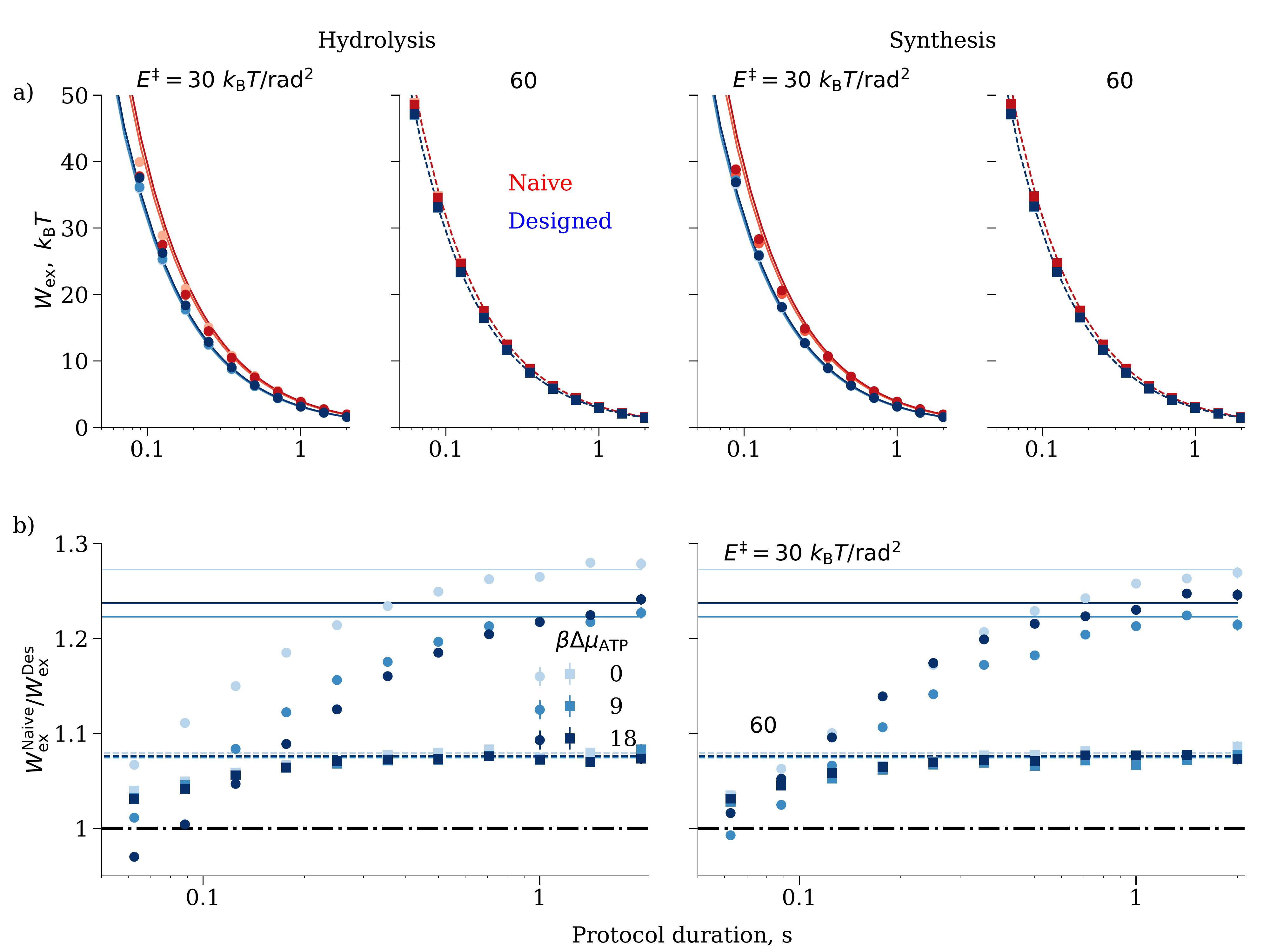}
    \caption{Excess work. (a) Excess work for naive (red) and designed (blue) protocols, and (b) ratio of naive and designed excess works, all as a function of protocol duration $\tDur$, for driven hydrolysis (left) and synthesis (right), for the full model (points) and the linear-response approximation (curves), for trap strengths $E^\ddag = 30~k_{\rm B}T/{\rm rad^2}$ (circles, solid curves) and $60~k_{\rm B}T/{\rm rad^2}$ (squares, dashed curves). Color intensity increases with the chemical drive $\Delta\mu_{\rm ATP}$.}
    \label{fig:work-ratio-SI}
\end{figure*}

Figure~\ref{fig:work-ratio-SI} shows naive and designed excess works and their ratio, $W_{\rm ex}^{\rm Naive}/W_{\rm ex}^{\rm Des}$~\cite[App.~B]{Barrier}, during driven hydrolysis and synthesis, for two trap strengths and three chemical drives $\Delta \mu_{\rm ATP}$. 

Longer protocol durations keep the system closer to equilibrium, increasing the accuracy of the linear-response approximation. The designed protocol's excess work agrees with its linear-response approximation starting at shorter durations than the naive protocol's does (Fig.~\ref{fig:work-ratio-SI}a). 

For virtually all durations explored, designed protocols save energy compared to naive protocols (Fig.~\ref{fig:work-ratio-SI}b). The performance of designed protocols (relative to naive), as quantified by the ratio of designed to naive excess works, improves with protocol duration until it saturates at the linear-response approximation. This excess-work ratio decreases with trap strength, and is qualitatively similar in both hydrolysis and synthesis directions. 

\section{Simulation methods}
\subsection{Torque autocovariance}
For $\Gamma \to \infty$, we discretize the Langevin equation~\eqref{tot-lan-dyn} to first order in $\Delta t$ and compute $\theta_\ell$ ($1\leq \ell\leq \tDur/\Delta t$) using 
\begin{align}
    \theta_{\ell} = \theta_{\ell-1} + \beta D 
    \torq^{\rm tot}_{\rm PMF} 
    (\theta_{\ell-1}|\lambda)~\Delta t + \sqrt{2D\, \Delta t}~\tilde{\eta}_{\ell-1} \ , \label{des-LE}
\end{align}
where $\torq^{\rm tot}_{\rm PMF}(\theta|\lambda)\equiv -\partial_\theta U^{\rm tot}_{\rm PMF}(\theta|\lambda)$ is the sum of torques derived from the underlying PMF and trap potential. $\tilde{\eta}_\ell$ is a normal random variable with zero mean and unit variance, and $\theta_0$ is drawn from the canonical distribution, i.e., proportional to $e^{-\beta U^{\rm tot}_{\rm PMF}(\theta|\lambda)}$ for $\theta \in[-\pi/2 +2\pi\lambda,\pi/2 +2\pi\lambda]$. (Notice that the trap potential $U_{\rm trap}(\theta|\lambda)$ has two minima separated by $\pi$~rad, and we initialize the $\gamma$-shaft's location in the well whose minimum is at $2\pi\lambda$~rad, where $\lambda \in [0,1]$.) For each fixed trap minimum $\lambda$, we generate a time-series of length $500$~s of the external torque $\torq_{\ell}\equiv-\partial_\lambda U_{\rm trap}(\theta_{\ell}|\lambda) = 2\pi E^{\ddag} \sin~2(\theta_{\ell}-2\pi\lambda)$ (see main text). We discard the initial 0.25 s (roughly 50$\times$ the torque relaxation time) of each trajectory, and use the remaining part to compute the torque autocovariance function $\langle \delta \tau_0~\delta \tau_{\ell} \rangle_\lambda$. The torque autocovariance in Fig.~\ref{fig:fric-var-relax}b is an average over three such trajectories initiated from three different initial conditions $\theta_0$.

For finite $\Gamma$, we again confine the bead attached to the $\gamma$-shaft in the sinusoidal potential $U_{\rm trap}(\theta|\lambda)$ [Eq.~\eqref{trap}], so that the system evolves according to~\eqref{W-dynamics-main} and switches its effective potential $U_n$ following the forward~\eqref{fwdR} and backward rates~\eqref{bwdR}. We follow the same numerical simulation procedure as discussed above for $\Gamma \to ~\infty$ to compute the friction coefficient and related quantities displayed in Fig.~\ref{fig:friction-SI}.

\subsection{Time-integrated average flux} 
\label{j-bar}
During a protocol, the system in general lags behind the trap's minimum, with magnitude depending on the protocol duration. To quantify rotation of the $\gamma$-shaft during a protocol completing a full rotation in a given duration $\tDur$, we numerically compute the time-integrated average flux
\begin{align}
    \bar{J} \equiv \dfrac{1}{2\pi} \int_0^{\tDur}~{\md}t~\int_0^{2\pi}~{\md}\theta~J(\theta,t) \ ,
\end{align}
for probability flux $J(\theta,t)$. $\bar{J} = 1$ corresponds to a complete $\gamma$-shaft rotation over the course of the respective protocol.

For all observed protocol durations and both protocols, $\bar J$ eventually approaches unity (see Sec.~``Simulation parameters"), indicating that the system reaches a nonequilibrium steady state.

\subsection{Escape fraction}
The external trap has two identical energy wells separated by 180$^\circ$. When the protocol duration is sufficiently short (so driving is sufficiently rapid), some trajectories cross the intervening barrier to the adjacent well. Our theoretical treatment assumes that the trap is sufficiently strong to prevent any such fluctuation, so we discard these trajectories.

Numerical simulations of the full model show that for $\Gamma\to\infty$ this escape fraction decreases (increases) with the chemical drive $\Delta \mu_{\rm ATP}$ when dynamically driving hydrolysis (synthesis), since the system moves with (against) the potential gradient due to $U_{\rm PMF}(\theta)$. Similarly, for finite $\Gamma$, larger $\Delta \mu_{\rm ATP}$ favors the $\gamma$-shaft's rotation in the hydrolysis direction~\eqref{W-eqn} reducing the proportion of trajectories that escape from the local well.

For $\Gamma\to \infty$, in the fastest simulated protocols dynamically driving hydrolysis (synthesis) and for the lowest (highest) $\Delta \mu_{\rm ATP}$, we find that $0.005\%$ ($0.5\%$) of trajectories escape the local trap. The fraction decreases with protocol duration, as expected. For the stronger trap ($E^\ddag = 60~k_{\rm B}T/{\rm rad^2}$), no trajectories escape across all observed durations.

\subsection{Excess work} 
To numerically compute the excess work, we first discretize each protocol for a given protocol duration $\tDur$. We compute the external work during a single trajectory by evaluating the net change in internal energy due to the control-parameter change from an initial value $\lambda_{\rm i}=0$ to $\lambda_{\rm f}=1$ in $\tDur$:
\begin{align}
    w = \sum_{\ell=0}^{\tDur/\Delta t-1}~[U_{\rm trap}(\theta_{\ell},\lambda_{\ell+1})-U_{\rm trap}(\theta_{\ell}, \lambda_{\ell})] \ ,
\end{align}
where $\theta_\ell$ is obtained from the discretized version of the Langevin equation at fixed $\lambda_\ell$, \eqref{des-LE} for $\Gamma \to \infty$ and analogously for finite $\Gamma$. When driving hydrolysis, we advance $\lambda$ from $\lambda_{\rm i}$ to $\lambda_{\rm f}$, whereas when driving synthesis, we reduce it from $\lambda_{\rm f}$ to $\lambda_{\rm i}$. The excess work is the work minus the equilibrium free-energy difference: $w_{\rm ex} \equiv w - \Delta F$. Here $\Delta F = \mp 3 \Delta \mu_{\rm ATP}$, where the `-' (`+') sign corresponds to  hydrolysis (synthesis), and the factor of 3 represents the 3 ATP hydrolyzed or synthesized per complete rotation. Finally, we average over the $\mathcal{N}_{\rm R}$ realizations that stay in the trap's local well (see `Escape fraction' section) to get the excess work $W_{\rm ex} \equiv \langle w_{\rm ex} \rangle$. To exclude transient relaxation to the nonequilibrium steady state, we compute the excess work starting after two full rotations (Figs.~\ref{fig:W-fig-work} and \ref{fig:work-hydro-W}) or four full rotations (Fig.~\ref{fig:work-ratio-SI}). Similar to the procedure in this section for excess work, we compute time-integrated average flux, escape fraction, and excess power in the numerical simulations.

\subsection{Simulation parameters}
\label{sim-para}
In our model, the system relaxation time in the trap is (to a harmonic approximation) $t_{\rm trap} \equiv \dfrac{k_{\rm B}T}{D E^\ddag}= 1.2-2.4 \times 10^{-3}$ s for trap strengths $E^\ddag = 30~k_{\rm B}T/{\rm rad}^2$ and  $60~k_{\rm B}T/{\rm rad}^2$, and diffusion constant $D=13.7$~rad$^2$/s. Therefore, for each numerical simulation we generally choose discretization timestep $\Delta t = 10^{-5}$ s, at least two orders of magnitude shorter than the relaxation time. We average over 3 independent trajectories to obtain the friction coefficient and related observables in Figs.~\ref{fig:fric-var-relax}(b-g) and \ref{fig:friction-SI}. 

For computing excess work, time-integrated average flux, and escape fraction when $\Gamma \to \infty$, we rotate the $\gamma$-shaft with protocol durations as fast as $\tDur = 2^{-4}$~s (i.e., 62.5 ms), $\approx$12.5$\times$ slower than the maximum torque relaxation time ($\sim 5$~ms). The data shown in Figs.~\ref{fig:ex-pow-main}, \ref{fig:ex-pow-SI}, and \ref{fig:work-ratio-SI} are obtained by averaging over $10^{5}$ trajectories during a complete rotation of the $\gamma$-shaft following four complete rotations to allow the system to reach steady state.

For the computation of excess work for finite $\Gamma$ (Figs.~\ref{fig:W-fig-work} and \ref{fig:work-hydro-W}), we rotate the $\gamma$-shaft in the range of protocol durations $2^{-1}-2^{4}$~s, where the fastest protocol has duration $\approx$5$\times$ the maximum of the torque relaxation time for the slowest potential-switching rate ($\Gamma=10~{\rm s}^{-1}$). We average over $10^{3}$ trajectories during a complete rotation of the $\gamma$-shaft following two complete rotations to allow the system to reach steady state. In order to increase the accuracy of the Langevin simulations, we used the discretization time $\Delta t = 10^{-6}$~s. 

For the largest finite $\Gamma$ and longest protocol durations, this timestep still produces discretization artifacts (seen in the modest disagreement between the full model and the linear-response approximation in Figs.~\ref{fig:W-fig-work} and \ref{fig:work-hydro-W}), but an even shorter timestep is computationally prohibitive.

\providecommand{\latin}[1]{#1}
\makeatletter
\providecommand{\doi}
  {\begingroup\let\do\@makeother\dospecials
  \catcode`\{=1 \catcode`\}=2 \doi@aux}
\providecommand{\doi@aux}[1]{\endgroup\texttt{#1}}
\makeatother
\providecommand*\mcitethebibliography{\thebibliography}
\csname @ifundefined\endcsname{endmcitethebibliography}
  {\let\endmcitethebibliography\endthebibliography}{}

\end{document}